\documentclass[prd,twocolumn,showpacs,nofootinbib]{revtex4}
\usepackage{times}
\usepackage{natbib}
\usepackage{epsfig}

\newcommand{\up}[1]{{\rm #1}}
\newcommand{\xiobs}{\xi_\up{obs}}
\newcommand{\xig}{\xi_\up{obs}(\sigma,\pi)}
\newcommand{\xint}{\xi_{gg}}
\newcommand{\xiz}{\xi_{zz}}
\newcommand{\xlens}{\xi_\up{lens}}
\newcommand{\ximb}{\xi_{ll}}
\newcommand{\xic}{\xi_{gl}}
\newcommand{\xiB}{\Delta\xi_\up{BAO}}
\newcommand{\sigB}{\sigma_\up{BAO}}
\newcommand{\xilin}{\xi_{mm}}
\newcommand{\plin}{P_{mm}}

\newcommand{\bdv}[1]{{\bf #1}}
\newcommand{\kms}{\, {\rm km}\, {\rm s}^{-1}}
\newcommand{\mpc}{\, {\rm Mpc}}
\newcommand{\hmpc}{{h^{-1}\mpc}}
\newcommand{\OM}{\Omega_m}
\newcommand{\rc}{c_{gm}}
\newcommand{\rbao}{r_\up{BAO}}

\newcommand{\DS}{\Delta\Sigma}

\begin{document}

\title{Gravitational Lensing Effects on the Baryonic Acoustic Oscillation
Signature in the Redshift-Space Correlation Function}

\author{Jaiyul Yoo$^1$}
\altaffiliation{jyoo@cfa.harvard.edu} 
\author{Jordi Miralda-Escud\'e$^{2,3}$}
\affiliation{$^1$Harvard-Smithsonian Center for Astrophysics,
Harvard University, 60~Garden Street, Cambridge, MA 02138}
\affiliation{$^2$Instituci\'o Catalana de Recerca i Estudis Avan\c cats,
Barcelona, Catalonia}
\affiliation{$^3$Institut de Ci\`encies del Cosmos, Universitat
de Barcelona / IEEC, Barcelona, Catalonia}

\begin{abstract}
Measurements of the baryonic acoustic oscillation (BAO) peak in the 
redshift-space correlation function yield the angular diameter distance
$D_A(z)$ and the Hubble parameter $H(z)$ as a function of redshift,
constraining the properties of dark energy and space curvature. We
discuss the perturbations introduced in the galaxy correlation function
by gravitational lensing through the effect of magnification bias and
its cross-correlation with the galaxy density. At the BAO scale,
gravitational lensing
adds a small and slowly varying component to the galaxy correlation function
and does not change its {\it shape} significantly, through which
the BAO peak is measured. The relative shift in the
position of the BAO peak caused by gravitational lensing in the
angle-averaged correlation function is $10^{-4}$ at $z=1$, rising to
$10^{-3}$ at $z=2.5$. Lensing effects are stronger near the
line-of-sight, however the relative peak shift increases only to
$10^{-3.3}$ and $10^{-2.4}$ at $z=1$ and $z=2.5$, when the galaxy
correlation is averaged within 5 degrees of the line-of-sight
(containing only 0.4\% of the galaxy pairs in a survey).
Furthermore, the lensing contribution
can be measured separately and subtracted from the observed correlation
at the BAO scale.
\end{abstract}


\maketitle

\section{INTRODUCTION}
\label{sec:intro}
A fundamental probe to the nature of the accelerated 
expansion of the universe is the comoving distance
corresponding to a redshift interval, $d\chi = dz / H(z)$, where $H(z)$
is the Hubble constant at redshift $z$.
The integrated function is related to the angular
diameter distance, $D_A(z)= \chi(z)/(1+z)$ for a flat model.
Deviations from this relation between $d\chi/dz$ and $D_A(z)$ are a
probe to space curvature, so far consistent with zero \citep{KODUET08}.
Recently, particular attention is being paid to
baryonic acoustic oscillations (BAO) in galaxy two-point statistics, 
as they provide a known physical scale tied to the sound horizon at the 
baryon decoupling epoch. 
Measurements of the BAO scale in the galaxy
correlation function can be used to infer both $H(z)$ and $D_A(z)$
(see, e.g., \cite{SEEI03,DETF06,
EIBLET05,TEEIST06,PENIET07,PASCET07,OKMAET08} and
see also, \cite{EISEN05} for their sensitivity to cosmological parameters).

  Gravitational lensing introduces perturbations on the galaxy
correlation function by deflecting light rays from galaxies 
(see, e.g., \cite{GUNN67,BLNA86,KAISE92}).
The main effect arises from the lensing magnification of the sky
area and the flux of each galaxy, known as magnification bias
\cite{TURNE80,NARAY89}. This results
in additional contributions to the observed galaxy correlation as a
function of separation in redshift-space 
\cite{MATSU00,VADOET07,YOO09}.
Another effect, which we shall not consider here, is the smoothing of
the BAO peak caused by changes in the observed angular separation of
galaxy pairs due to the lensing deflection, which induces a negligibly
small shift on the position of the BAO peak (e.g., \cite{VADOET07}).

We examine the modifications of the observed galaxy two-point correlation
function $\xig$ in redshift-space
due to gravitational lensing, where 
$\sigma=D_A(z)(1+z)\phi$ and $\pi=\Delta z/H(z)$ are the comoving 
separations of galaxy pairs across and along the line-of-sight in 
redshift-space, and $\phi$ and $\Delta z$ are the observable
angular and redshift separations. We evaluate the magnitude of the
lensing contribution to clarify the level of accuracy at which the
gravitational lensing effect needs to be taken into account for
precision measurements of the BAO scale. We show that 
despite previous claims to the contrary \citet*{HUGALO07}
the effect of gravitational lensing is generally small for currently
planned surveys, because gravitational lensing hardly changes the correlation
function shape at the BAO scale and in practice galaxy pairs are averaged
over a finite angular bin.
We adopt a flat $\Lambda$CDM cosmology with $\Omega_m=0.28$ and
$H_0 =70 \kms\mpc^{-1}$, according to recent
measurements of the cosmic microwave background \cite{KODUET08}.
We set the speed of light $c\equiv1$.

\section{Formalism}

  We first summarize the basic equations for computing galaxy two-point 
correlation functions. In the linear approximation,
the intrinsic galaxy correlation function is $\xint(r)=b^2 \xilin(r)$,
where $b$ is a constant linear bias factor and $\xilin(r)$ is the mass
correlation function. The redshift-space galaxy correlation function
is computed by Fourier transforming the linearly biased matter
power spectrum $b^2\plin(k)$ with the redshift-space enhancement 
factor arising from peculiar velocities \cite{KAISE87},
\begin{equation}
\xiz(\sigma,\pi)=\int{d^3\bdv{k}\over(2\pi)^3}~e^{i\bdv{k}\cdot\bdv{s}}~
b^2\plin(k)~\left(1+\beta\mu_k^2\right)^2,
\end{equation}
where $\bdv{s}=(\sigma,\pi)$, $\mu_k=k_z/k$, $\beta=f/b$, 
$f=d\ln D/d\ln a$, and $D(z)$ is a growth factor of the matter density.
We use the \citet{SMPEET03} approximation for computing the non-linear
$\xilin(r)$ and $\plin(k)$.

Lensing
introduces two terms in the correlation function of galaxies above some
luminosity $L$. The first is
due to the auto-correlation of the magnification bias on two sources
at $z_1$ and $z_2$ ($z_1<z_2$),
\begin{equation}
\ximb(\sigma)=\left( 3H_0^2 \OM\alpha \right)^2
\int_0^{\chi_1}d\chi\left[{\chi (\chi_1-\chi)\over a(\chi)\chi_1}\right]^2 
w_p(\chi\phi) ~,
\label{eq:ximb}
\end{equation}
where $\alpha = - d\log \bar n_g/d\log L - 1$, and $\bar n_g(L,\bar z)$
is the cumulative number density of galaxies with luminosity above $L$
at the mean source redshift $\bar z$.
We assume the two sources are at nearly the same redshift,
with a separation $\pi \ll \chi_1$. 
The dependence of the magnification bias on $\alpha$ arises from the
combination of the magnification of the sky area and the flux
amplification of the sources (see \cite{NARAY89,SCMEET05}).
The projected mass correlation function is
\begin{equation}
w_p(\sigma)=\int_{-\infty}^\infty d\pi~
\xilin\left( r=\sqrt{\sigma^2+\pi^2}\right) ~.
\label{eq:wp}
\end{equation}
The other term that is added to the observed galaxy correlation is due
to the cross-correlation of the intrinsic galaxy fluctuation and the
magnification bias. Since the matter fluctuation along the line-of-sight
is responsible for the magnification bias in the background galaxy, 
it correlates with the galaxy fluctuation and this cross-correlation is
\begin{equation}
\xic(\sigma,\pi) = 3H_0^2\OM\alpha \left[\int_0^{\chi_2}\!\!\!d\chi~
{\chi~(\chi_2-\chi)\over a(\chi)\chi_2}~
\xi_{gm}(r_1) ~+~ (1\leftrightarrow2) \right]~,
\label{eq:xic}
\end{equation}
where $r_1=\sqrt{\phi^2\chi_1^2+(\chi_1-\chi)^2}$, 
$r_2=\sqrt{\phi^2\chi^2+(\chi_2-\chi)^2}$,
$\bar\chi=(\chi_1+\chi_2)/2$, $\sigma = \phi\bar\chi$, $\pi=\chi_2-\chi_1$,
the galaxy-mass cross-correlation is $\xi_{gm}(r)=b\rc \xilin(r)$, and
$\rc$ is a galaxy-mass cross-correlation coefficient
(e.g., \cite{PEN98}). The two added terms exchanging the subindexes
(1,2) account for the effect of magnification bias in the background and
foreground galaxy, respectively.
After some rearrangement, we obtain,
in the approximation $\pi \ll \bar\chi$,
\begin{eqnarray}
\label{eq:xico}
\xic(\sigma,\pi) &=& 3H_0^2\OM\alpha (1+\bar z) ~ \\
 & \times & \left[ \pi~ w_{p,gm}(\sigma) + 2\int_{\pi}^{\infty} d\tau~ (\tau-\pi)
 ~\xi_{gm}(r) \right] ~,
\nonumber
\end{eqnarray}
where $w_{p,gm}$ is the same projected correlation function as in 
Eq.~(\ref{eq:wp}) for $\xi_{gm}$, and $r=\sqrt{\sigma^2+\tau^2}$. 
Equation~(\ref{eq:xico})
has been derived before without the inclusion of the second
term (e.g., \cite{BARTE95}), an approximation that is valid only when
$\sigma \ll \pi$, in addition to $\pi \ll \bar\chi$. This second term
is important for determining the functional shape of $\xi_{gl}$ over all
the redshift space, but is small in the region where the lensing effect
is strongest, at $\sigma \ll \pi$.
For the results presented here, we use the more exact
Eq.~(\ref{eq:xic}) for computing $\xic$.
The total, observed galaxy correlation function is
$\xig=\xiz(\sigma,\pi)+\ximb(\sigma)+\xic(\sigma, \pi)$.

\begin{figure}[t]
\centerline{\psfig{file=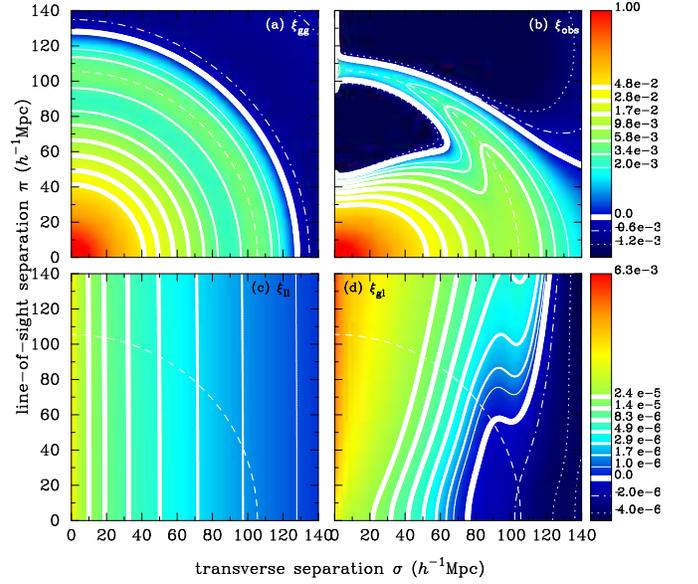, width=3.4in}}
\caption{(color online) Two-point correlation functions in 
redshift-space at $\bar z=0.35$. (a)~Intrinsic galaxy correlation function
$\xint$.  (b)~Observed galaxy correlation function
$\xiobs=\xiz+\ximb+\xic$. 
(c)~Magnification bias correlation function $\ximb$.
(d)~Cross-correlation function $\xic$ of the intrinsic galaxy
fluctuation and the magnification bias.
The color scale is proportional to the logarithm of the
correlation function at $\xi\geq1\times10^{-4}$ in the
top panels, and at $\xi\geq8\times10^{-7}$ in the bottom panels, below which
the scale is linear with $\xi$. White contours of different thickness
are as indicated in the right bars, with the thickest contour corresponding
to $\xi=0$. Negative contours are shown as dot-dashed and dotted curves.
Since the lensing effect is small, the redshift-space correlation function
$\xiz$ is similar to $\xiobs$ in Panel~($b$), except for the small spot
produced near $\sigma=0$, $\pi = \rbao$.
A galaxy bias factor $b=2$ and luminosity function slope $\alpha=2$
are assumed. The baryonic acoustic oscillation scale $\rbao$
is shown as a dashed circle for reference.}
\label{fig:four}
\end{figure}

\begin{figure}[t]
\centerline{\psfig{file=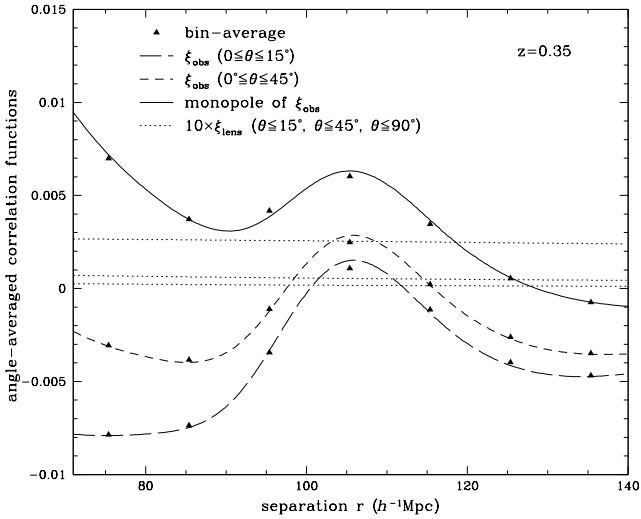, width=3.4in}}
\caption{Angle-averaged correlation functions and lensing contributions
at $\bar z=0.35$. Observed correlation function $\xig$ is averaged over
$0\leq\theta < 15^\circ$ (long dashed), $0\leq\theta < 45^\circ$
(short dashed), and over all angles (solid, monopole).
Lensing contribution ($\xlens=\ximb+\xic$) is shown multiplied by~10 and
averaged over the same angular intervals (dotted, from top to
bottom). Triangles show $\xiobs$ averaged over each radial bin of width
$10\hmpc$. }
\label{fig:bin}
\end{figure}

\begin{figure}[t]
\centerline{\psfig{file=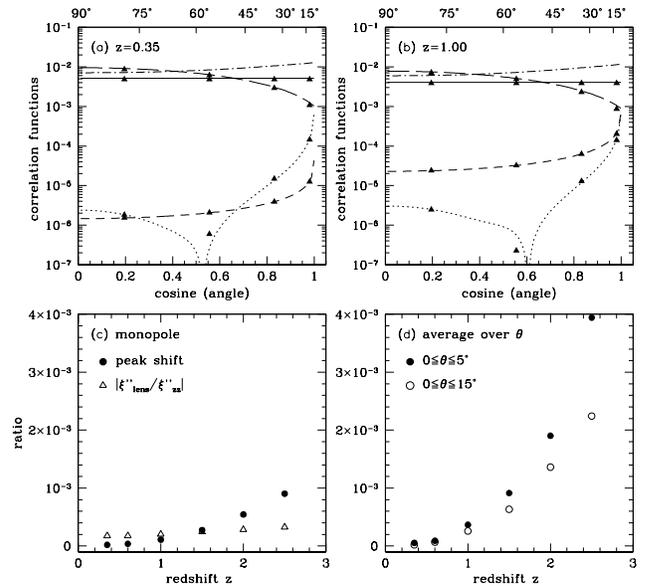, width=3.4in}}
\caption{Gravitational lensing effect on the correlation function
at the BAO scale. 
Upper panels: Intrinsic galaxy correlation ($\xint$, solid),
redshift-space correlation ($\xiz$, long dashed), magnification
bias correlation ($\ximb$, short dashed), and galaxy-magnification
cross-correlation ($\xic$, dotted) as a function of
cosine angle $\mu=\cos\theta$.
Note $\xic<0$ at $\mu\lesssim0.6$, where its absolute value is plotted.
Additional dot dashed curves show the BAO peak height $\xiB$, defined
in Eq.~(\ref{eq:xiB}). Triangles show correlation 
functions averaged over radial width $10\hmpc$ and 
angular width $22.5^\circ$. 
Bottom panels:
 As a function of redshift, circles and triangles represent
lensing contribution to BAO peak position shift and height
(Eqs.~(\ref{eq:xis}) and (\ref{eq:xiB})) averaged over all
angles (left), and filled and empty circles show lensing
contributions to BAO peak position shift averaged over
angles within 5 and 15 degrees (right).
We compute $\ximb$ and $\xic$ at $\mu\leq0.9999$ (corresponding to
$\sigma=1.5\hmpc$ at $r=\rbao$), beyond which the linear bias approximation
may be inaccurate (cf. Fig.~\ref{fig:ds}). }
\label{fig:ang}
\end{figure}

\begin{figure}[t]
\centerline{\psfig{file=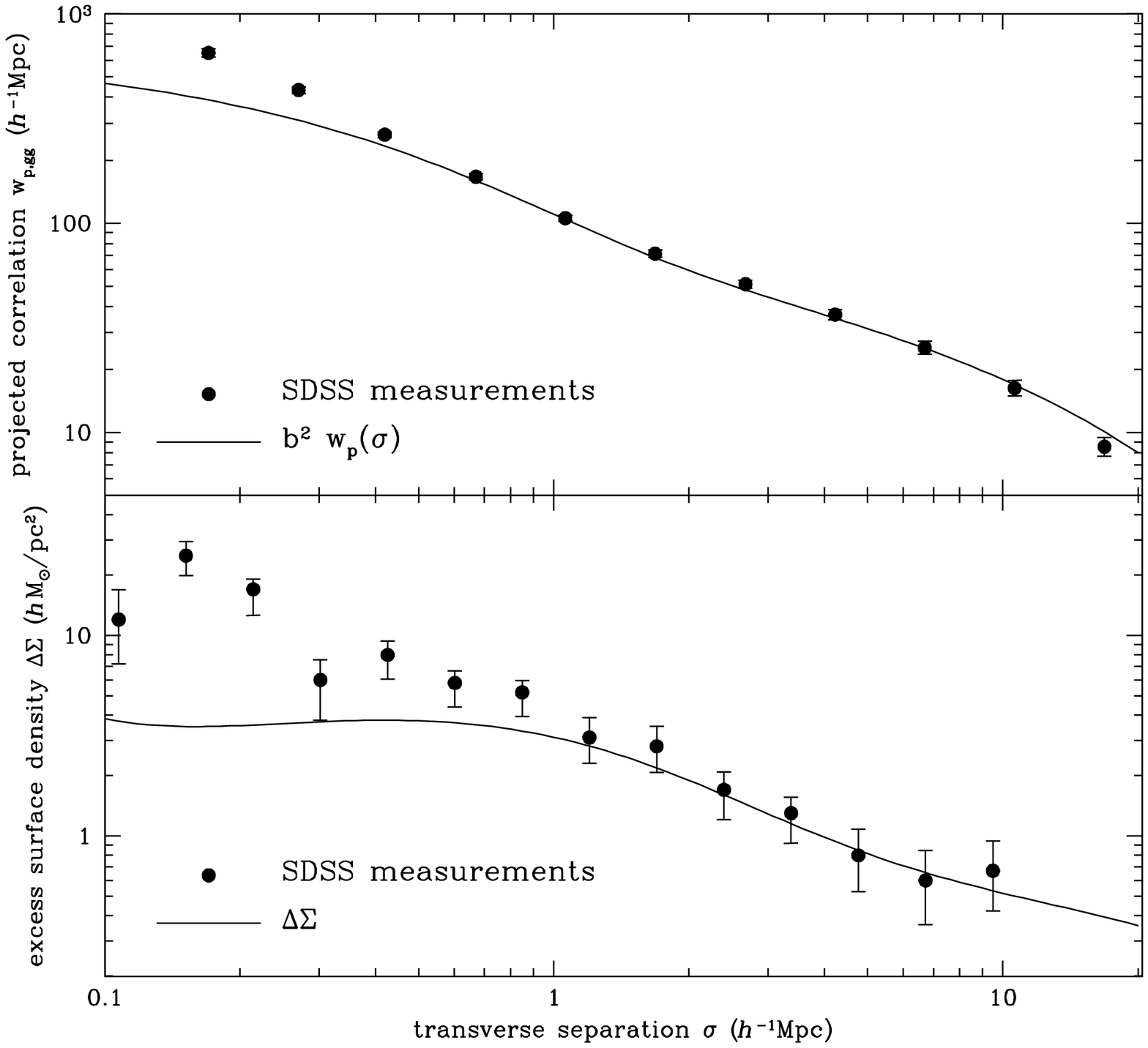, width=3.4in}}
\caption{Projected galaxy correlation $w_{p,gg}$ and excess surface density
$\Delta \Sigma$ computed from the non-linear mass correlation function
$\xi_{mm}$ (solid), compared with projected galaxy correlation
and lensing shear measurements from SDSS \cite{ZEZE05,SHJO04}. This
validates our modeling of lensing effects based on linear bias
for $\sigma \geq 1 \hmpc$.}
\label{fig:ds}
\end{figure}

\section{Results}

  Figure~\ref{fig:four} shows the two-point correlation functions in
redshift-space at $\bar z=0.35$. The upper
panels show the intrinsic galaxy correlation function $\xint$
(left) and the observed galaxy correlation function $\xiobs$
(right). We choose a galaxy bias $b=2$ at $\bar z=0.35$ and
a cross-correlation coefficient $\rc=1$, as measurements suggest for 
Sloan Digital Sky Survey (SDSS) luminous red
galaxy (LRG) samples (see, e.g., \cite{EIBLET05,TEEIST06,SHJO04}). 
The galaxy bias at other redshifts is computed 
assuming galaxies move as test
particles responding to gravity in the linear regime, in which case
$b(z)-1=[b(z=0)-1]/D(z)$, where $D(z)$ is the growth
factor normalized to unity at $z=0$ \cite{FRY96}.
Note that $\xint$ scales as $b^2$, and $\xiobs$ has an additional change
of its contours with bias through the $\beta$ parameter.

  The BAO scale is defined as the distance traveled by a sound wave up
to the baryon decoupling (drag) epoch at time~$t_d$,
$\rbao=\int_0^{t_d}c_s~(1+z)~dt = 155 ~\up{Mpc}$,
where $c_s^2=1/3(1+R)$, $R$ is the baryon-photon ratio, and we use the
\cite{EIHU98} fitting formula for computing $t_d$ (see also, 
\cite{HU05,KODUET08}). We indicate the BAO scale as a short-dash
circle in Fig.~\ref{fig:four}. The bump in the correlation function at this scale
shown by the contours of $\xiobs$ is the signature to be used to
measure $\rbao/D_A(z)$ and $\rbao\, H(z)$. The redshift-space
distortion squashes the contours of $\xiobs$ along the line-of-sight and
changes the shape of the BAO peak at each angle in the $\sigma$-$\pi$
plane. The lensing effect is very small,
and so the contours of $\xiobs$ in Fig.~\ref{fig:four} are nearly identical to
the contours of $\xi_{zz}$, except for a slight difference very close to
the line-of-sight ($\sigma \ll \pi$), where the lensing effect is strongest.

  The bottom panels show the correlation of the magnification bias
$\ximb$ (left) and the cross-correlation of the magnification bias
and the intrinsic galaxy fluctuation $\xic$ (right). 
We use $\alpha=2$ for the magnification bias, which is approximately the
value for an LRG sample with $L > 3 L_*$, close to the
threshold for the SDSS \cite{EIBLET05,COEIET08}. 
Note that the contour scale is
smaller by a factor 100 than that in the upper panels. The function
$\ximb$ decreases with $\sigma$ and depends very weakly on $\pi$ through
$\chi_1=\bar\chi-\pi/2$ in Eq.~(\ref{eq:ximb}), 
whereas $\xic$ decreases with $\sigma$ and increases with $\pi$.
The correlation $\xic$ contains a weak BAO ripple when $\sigma$ is near
the BAO scale, arising from the integration in Eq.~(\ref{eq:wp})
when the edge of the BAO sphere is seen in projection along the
line-of-sight. The lensing correlations are of course largest near
the line-of-sight at $\sigma \ll \rbao$, where the BAO peak of 
$\xic$ is washed out by the integration.

  Figure~\ref{fig:bin} shows $\xiobs(r)$ and
$\xlens(r)=\ximb(r)+\xic(r)$ at $\bar z=0.35$, averaged over volume with
different angular intervals. The solid line is the monopole of
$\xiobs$. The short dashed and long dashed lines show $\xiobs$
averaged only over the angles $\theta < 45^\circ$ and
$\theta < 15^\circ$ from the line-of-sight, respectively. The lensing
contributions are indicated by the three dotted lines, averaged over
the same angle intervals, from bottom to top; these curves are
multiplied by~10 to enable visualization.
Even within the narrow
interval $\theta < 15^\circ$ (which contains only 3.4\% of the galaxy
pairs), the lensing contribution to $\xiobs$ is 
$\sim3\times10^{-4}$, while the contrast of the BAO peak is
$\Delta \xi \sim 0.01$. 
Note that the lensing contribution is dominated by $\xi_{gl}$, and
therefore it scales as $\alpha b c_{gm}$.
The lensing effect adds only a small component to $\xiobs$ that is very
slowly varying with $r$, and
cannot alter the shape of the BAO peak in any appreciable way. 
\citet{VADOET07} also reached the same conclusion that the magnification
bias on the BAO peak shift is negligible, although they compared the
intrinsic galaxy correlation function $\xi_{gg}$ with the lensing contribution
in the transverse direction ($\pi=0$, $\sigma\simeq\rbao$).

  Figure~\ref{fig:ang} examines the gravitational lensing effect at the
BAO scale, $r=\rbao$, as a function of the cosine angle
$\mu=\cos\theta=\pi/r$. Note that an equal amount of volume is available
to measure the correlation function per interval $d\mu$. The upper panels
show $\xint$ (solid), $\xiz$ (long dashed), $\ximb$
(short dashed), and $\xic$ (dotted), at $\bar z=0.35$ and 
$\bar z=1$,
with galaxy bias factor $b=2$ and $b=2.3$, respectively. A fifth curve
(dot dashed) shows the BAO peak amplitude $\xiB$, which we define
in the next paragraph. All the functions are evaluated at $r=\rbao$.
The slope of the luminosity function is fixed to $\alpha=2$.
Triangles show the averaged correlation function
over angular bins of width $\Delta r = 10 \hmpc$ and
$\Delta \theta = 22.5^\circ$.

  To understand the effect of lensing on the BAO peak, one should note
that the ability to measure the peak position $\rbao$ depends on the
{\it shape} and {\it height} of the BAO peak, rather than the specific
value of $\xiobs$ at $\rbao$. For example, near the line of sight
($\mu=1$), the redshift-space correlation function
$\xiz(\simeq\xiobs)$ happens to be very close to
zero at $r=\rbao$, so a small lensing contribution can change
$\xiobs(\rbao)$ by an increased factor.
However, this is totally irrelevant for measuring the
BAO peak position and for quantifying the importance of lensing.
We therefore choose a definition of the BAO
peak height $\xiB$ in terms of the second derivative of $\xiz$ at
$\rbao$: 
\begin{equation}
\xiB(\theta)=~-{\sigB^2\over2}~\xiz''(\rbao,\theta),
\label{eq:xiB}
\end{equation}
where the prime indicates a partial derivative with respect to $r$ at
fixed angle $\theta$, and $\sigB$ is a constant that represents the
width of the BAO peak and can be adjusted to fit the peak height,
$\xiB$. This definition is exact when $\xiz$ is approximated as a
linear component plus a Gaussian bump of width $\sigB/\sqrt{2}$ at
$r=\rbao$.
We choose $\sigB=15 \hmpc$, which results in
the dot dashed curves shown in Fig.~\ref{fig:ang}. We see that $\xiB(\mu)$
increases slightly with $\mu$, in contrast to $\xiz(\mu)$
which drops sharply with $\mu$ close to $\mu=1$
(the width of the BAO peak is actually narrower at $\mu\simeq1$ than for
the monopole, so $\xiB$ increases less with $\mu$ if this is taken into
account). This indicates that Eq.~(\ref{eq:xiB})
remains a very good approximation, 
as $\xi_{zz}$ has negligible curvature around the BAO scale once the Gaussian
component is removed.

  The ratio $\xlens/\xiB$ is $\lesssim 10^{-2.5}$ over most of the
volume at $\bar z<1$, and is $\sim$ 2\% at $\theta \leq 15^\circ$. 
At $\bar z > 1$,
the $\ximb$ lensing contribution becomes dominant and increases roughly
as $\bar\chi^3$. Since the lensing contribution to $\xiobs$ has a very
slow variation with $r$, the effect on the measurement of the BAO scale
is much smaller than $\xlens/\xiB$. The radial shift 
$\Delta r_\up{max}=\rbao-r_\up{obs}$
in the maximum of the correlation function at fixed $\theta$ is
\begin{equation}
  {\Delta r_\up{max} \over \rbao} =~
     - {\xlens' \over 2 \xiB}~{\sigB^2\over\rbao } ~.
\label{eq:xis}
\end{equation}
Note that the shift $\Delta r_\up{max}$ in Eq.~(\ref{eq:xis}) is
independent of our choice of the $\sigB$ value.
The bottom panels of Fig.~\ref{fig:ang} show this relative radial peak
shift (circle), and the relative change in the BAO peak height,
$|\xlens''/\xi''_{zz}|$ (triangle), for the angle-averaged case
(left), and averaging over $\theta\leq 5^\circ$ (right).
The peak shift $\Delta r_\up{max}/\rbao$ is, for the angle-averaged
case, $\sim 10^{-4}$ at $\bar z\leq1$, rising to $\sim 10^{-3}$ at
$z=2.5$. When restricted to the narrow region near the line-of-sight
$\theta \leq 5^\circ$, this peak shift increases by a factor of only
$\sim 4$, still remaining a very small effect. We have checked that
even at $1^\circ$ from the line-of-sight the peak shift due to lensing
grows only by another factor of 2 compared to the $\theta \leq 5^\circ$
case. 

Naturally, in any galaxy survey, the error to which the BAO peak
position can be measured in a region within an angle $\theta$ of the
line-of-sight is increased by at least the factor $\sqrt{2}/\theta$
compared to the angle-averaged measurement,
owing to the increased shot noise and sampling variance.
For the purpose of measuring the radial BAO peak position, the galaxy
correlation function always needs to be averaged over a finite angular bin,
and no substantial added precision is obtained for very small angles
from the line-of-sight.  Therefore,
lensing effects on the BAO peak position will always be very
small  in practice. The lensing contribution to the BAO height is
$\sim 2\times 10^{-4}$ for the monopole, increasing very slowly with
redshift, and is actually smaller near the line-of-sight. This shows
that even though the value of $\xlens$ at $\rbao$ is largest
near the line-of-sight, its effect on the BAO peak is not necessarily
so, because adding a constant to the correlation function is irrelevant
for measuring the BAO peak.

  The impact of gravitational lensing on the BAO peak was previously
discussed by \citet{HUGALO07}. We disagree
with their conclusion that there are large lensing effects.
\citet{HUGALO07}
define a fractional change in the BAO peak height as
$(\xiobs - \xint)/\xint$. As discussed above, this quantity is
irrelevant because adding a constant to the correlation function has
no effect on the measurement of the BAO peak. Moreover, the value of
$\xint$ at the BAO peak, or of $\xiz$ when the correlation is measured
in redshift space over a specific angular range, may happen to be near
zero, which may give rise to a large fractional change of $\xint$ due
to the lensing effect, but this is equally irrelevant: only the
amplitude of the BAO peak matters, and not the value of $\xi$ at the
peak. 

\citet{HUGALO07} also claim that lensing has strong effects in
the line-of-sight direction.\footnote{\citet{HUGALO07} calculate the 
line-of-sight galaxy-lensing correlation using the projected mass 
auto-correlation with a constant bias
factor extrapolated to zero separation. This yields the average lensing
convergence behind a random mass particle (times the assumed bias
factor), instead of the convergence behind the center of a galaxy. In
reality, whenever the lensing effect is observed exactly on the
line-of-sight to a galaxy with a central cusp, strong lensing must
occur.} 
In reality, the correlation
function can only be observed averaged within a finite angle of the
line-of-sight, and can only be computed using a constant bias down to
some minimum separation for which the linear bias approximation for the
projected galaxy-mass cross-correlation is reasonable. This explains
why \citet{HUGALO07} find a shift in the BAO peak location on the
line-of-sight direction of 3\%  that is nearly redshift independent
(see their Fig.~8a; our values of bias and slope correspond to
$(5s-2)/b=2$ in their notation), whereas we find that within 5 degrees
of the line-of-sight the shift increases rapidly with redshift and
reaches only 0.4\% at $z=2.5$, and within 1 degree of the line-of-sight
the shift is larger by only a factor $\sim 2$. For the
angle-averaged lensing effect, we also disagree with the results of
\citet{HUGALO07}: they find a peak shift of $0.4\%$ at z=2.5 (for the
same bias and slope we use), compared to our result of $0.1\%$.

We note that if one insists on measuring the correlation of galaxies
exactly on the line-of-sight, strong lensing occurs and the background
galaxy is imaged into an Einstein ring, an effect that is already
detected (see, e.g., the Sloan Lenses ACS Survey
\cite{TRKOET05}). However, this lensing effect has no special 
feature at the BAO scale and has no interesting effect on the ability
to measure the BAO peak in the galaxy correlation function.

  Finally, we comment on the way to observationally separate the lensing
contribution from $\xiobs$. Considering galaxies of two types with bias
factors $b_1$ and $b_2$ and luminosity function slopes $\alpha_1$ and
$\alpha_2$, the parity of the correlation functions $\xig$ is even under
a change of sign of $\pi$, except for the galaxy-magnification
cross-correlation $\xic$, which is different depending on the galaxy
type that is in the foreground or background. For simplicity, we
consider the case $\sigma \ll \pi$, when the second term in 
Eq.~(\ref{eq:xico}) can be neglected. Hence, the
asymmetry of the cross-correlation function of two different types of
galaxies yields the galaxy-lensing contribution:
\begin{eqnarray}
\xiobs(\sigma,\pi) &-& \xiobs(\sigma,-\pi) = \xic(z_1<z_2)-\xic(z_2<z_1) 
\nonumber \\
&=& (b_1 \alpha_2 - b_2 \alpha_1)\, 3H_0^2\OM ~(1+\bar z) ~ \pi \, w_p(\sigma)~.
\end{eqnarray}
Consequently, it is in principle possible to directly separate the
$\xic$ contribution at the BAO scale purely from observations. 
Alternatively, since the lensing contribution is very small at the BAO
scale, one can measure $\xic$ at large $\pi$ (e.g.,
\cite{SCMEET05}), where the contribution from $\xiz$ is small, and
use the known dependence on $\pi$ to subtract its contribution from the
measurements of $\xiobs$ at the BAO scale.

\section{Discussion}
\label{sec:dis}

  We have shown that modifications of the galaxy correlation function
caused by gravitational lensing are a tiny effect for the purpose of
measuring the BAO scale.
The lensing contribution to the correlation function near the BAO peak
is $\xlens \sim 10^{-4}$ at $\bar z < 1$, even within $15^\circ$ of
the line-of-sight.
Moreover, the lensing contribution is nearly constant as a function of
radius, so the ability to measure the BAO peak and its shape in any
galaxy survey is not affected. The galaxy correlation function is averaged
over a finite angular bin, further suppressing the lensing effect.
The shift in the
position of the BAO peak due to lensing in the angle-averaged
correlation function is less than 1 part in $10^4$ at $\bar z\leq 1$ and
it increases to $\sim 10^{-3}$ at $\bar z=2.5$. This peak shift is
increased by a factor of only 4 within 5 degrees of the line-of-sight,
where just 0.4\% of the galaxy pairs are available for measuring the
correlation function. The lensing effect increases with the luminosity
function slope $\alpha$, but not sufficiently to make it substantial for
any known population of sources.

As we discussed in Sec.~III, when two types of galaxies are used to measure
the correlation function, we can directly measure the lensing contribution
$\xlens$ from observations and subtract it before we estimate the BAO peak
position.
  In general, the addition of any broadband power to the correlation
function by known physical effects can be modeled and removed. The
method for measuring the position of the BAO peak may be optimized to
minimize the dependence on added broad-band power from several physical
effects in addition to lensing \cite{SESIET08,XUWHET10}. Therefore, the lensing
effect we have computed here is likely to be further reduced when using
optimized definitions of the BAO scale.

  The linear bias approximation we have used here becomes invalid for
computing the galaxy-magnification cross-correlation
in Eq.~(\ref{eq:xic}) close to the line-of-sight, when the
transverse separation $\sigma$ is small. The bias coefficients $b$ and
$\rc$ may be scale-dependent, and other non-linear terms may become
important. However, the correlations
induced by lensing can be tested by independent observations of
lensing effects around galaxies \cite{SHJO04,MHSGET05,MSCBHB06}.
Figure~\ref{fig:ds} shows
the projected galaxy correlation function $w_{p,gg}$ \cite{ZEZE05}
and the excess surface density $\DS$ inferred from weak lensing measurements
\cite{SHJO04}, for the SDSS main sample of galaxies. 
Also shown as solid lines are the result for $w_p$ from the
mass correlation function used in this paper 
and for the excess surface density,
\begin{equation}
\DS(\sigma) \propto {2\over\sigma^2}
 \int_0^\sigma w_{p,gm}(R) ~R ~dR - w_{p,gm}(\sigma) ~.
\end{equation}
Normalization is adjusted to match the data on large scales. For
other types of galaxies one can use the results of 
\citet{SHJO04} to match the required value of $b\rc$.

  The measurements are in reasonable agreement with the linear bias
approximation at $\sigma\geq1\hmpc$. At smaller separations, the shape
of the galaxy-mass cross-correlation is clearly steeper than our simple
model. This is not surprising because galaxies tend to occupy the
central positions in halos. The mass auto-correlation function at these
small scales reflects the density profiles of dark matter halos, which
have a slope that gradually flattens at small radius, whereas galaxies
are more centrally concentrated than mass in halos (see, e.g.,
\cite{ZEWE04,YTWZKD06} for the one-halo contributions). 
These small scales would affect the BAO signal at angles
$\theta \lesssim 0.5^\circ$ for the SDSS main galaxy samples
and $\theta\lesssim1.0^\circ$ for the LRG samples (see, e.g., 
\cite{MHSGET05}), containing a very small fraction of the
galaxy pairs. We conclude that non-linear effects can be neglected,
except within angles as small as $1.0^\circ$, where they can be
calibrated to the observational results.

\acknowledgements
We thank Adam~Lidz, Daniel~Eisenstein, David Weinberg, and Uro{\v s} 
Seljak for useful discussions.
 J.~Y. is supported by the Harvard College Observatory under the
Donald~H. Menzel fund.  J.~M. is supported by the Spanish grants
AYA2006-06341, AYA2006-15623-C02-01, AYA2009-09745 and MEC-CSD2007-00060.

\vfill

\bibliography{ms.bbl}

\end{document}